\documentstyle[prd,aps,preprint,tighten,epsfig]{revtex}

\begin{document}

\draft
\preprint{\begin{tabular}{r}
{\bf hep-ph/0009294} \\
{LMU-00-14} \\
{~}
\end{tabular}}

\title{Commutators of Lepton Mass Matrices, CP Violation, and \\
Matter Effects in Medium-Baseline Neutrino Experiments}
\author{\bf Zhi-zhong Xing}
\address{Theory Division, Institute of High Energy Physics, P.O. Box 918, 
Beijing 100039, China \\
and \\
Sektion Physik, Universit$\it\ddot{a}$t M$\it\ddot{u}$nchen,
Theresienstrasse 37A, 80333 M$\it\ddot{u}$nchen, Germany \\
({\it Electronic address: xing@theorie.physik.uni-muenchen.de}) }
\maketitle

\begin{abstract}
We introduce the commutators of lepton mass matrices to describe
the phenomenon of lepton flavor mixing, and establish their relations
to the effective Hamiltonians responsible for the propagation of
neutrinos. The determinants of
those commutators are invariant under matter effects, leading to an 
instructive relationship between the universal CP-violating parameters 
in vacuum and in matter.
In the scenario of low-energy (100 MeV $\leq E \leq$ 1 GeV) and 
medium-baseline (100 km $\leq L \leq$ 400 km) neutrino
experiments, we illustrate the features of lepton flavor
mixing and CP violation. The terrestrial matter effects on CP- and
T-violating asymmetries in $\nu_e \leftrightarrow \nu_\mu$ and
$\overline{\nu}_e \leftrightarrow \overline{\nu}_\mu$ 
neutrino oscillations are also discussed. We demonstrate that
a relatively pure signal of leptonic CP violation at the percent level
can be established from such medium-baseline experiments with 
low-energy neutrino beams.

\end{abstract}



{\Large\bf 1} ~
Robust evidence for the long-standing anomalies of solar and
atmospheric neutrinos has recently been reported by the 
Super-Kamiokande Collaboration \cite{SK98}. It strongly implies
that neutrinos are massive and lepton flavors are mixed.
The admixture of three different lepton flavors generally
involves non-trivial complex phases, leading to the phenomenon
of CP violation \cite{Cabibbo}. Leptonic CP violation can manifest itself 
in neutrino oscillations. In reality, to measure CP-violating 
asymmetries needs a new generation of accelerator neutrino experiments 
with very long baselines \cite{LB}. In such long-baseline experiments 
the earth-induced matter effects, which are likely to 
deform the neutrino oscillation patterns in vacuum and to fake
the genuine CP-violating signals, must be taken into account.

Recently some attention has been paid to an interesting possibility
to measure lepton flavor mixing and CP violation
in the {\bf medium}-baseline neutrino experiments with
{\bf low}-energy beams \cite{Richter,Minakata,Sato}. The essential
idea is on the one hand to minimize the terrestrial matter effects, 
which are more significant in the long-baseline
neutrino experiments, and on the other hand to 
obtain the fast knowledge of neutrino mixing and CP violation
long before a neutrino factory based on the muon storage ring
is really built. Although such low-energy and medium-baseline 
neutrino experiments may somehow suffer from the problems such
as smaller detection crosss sections and larger beam opening angles,
they can well be realized by choosing the optimum baseline length and 
beam energy. 
They are also expected to be complementary to the high-energy and 
long-baseline neutrino experiments, towards a full 
determination of lepton flavor mixing and CP-violating parameters.

The purpose of this paper is three-fold. First, the commutators of
lepton mass matrices are introduced to describe the phenomenon 
of lepton flavor mixing, and their relations to the effective
Hamiltonian responsible for the propagation of Dirac and Majorana
neutrinos are established. An elegant relationship 
between the universal CP-violating parameters in matter and in vacuum,
no matter whether neutrinos are Dirac or Majorana particles, can
then be derived from the determinants of those commutators, which
are invariant under matter effects.
Secondly, we study the features of lepton flavor mixing 
and CP violation in the scenario of low-energy 
(100 MeV $\leq E \leq$ 1 GeV) and medium-baseline 
(100 km $\leq L \leq$ 400 km) neutrino experiments. The
terrestrial matter effects on the elements of the flavor mixing matrix 
and the rephasing-invariant measure of CP violation are in
particular illustrated. Finally we analyze the CP- and T-violating
asymmetries in $\nu_e \leftrightarrow \nu_\mu$ and
$\overline{\nu}_e \leftrightarrow \overline{\nu}_\mu$
neutrino oscillations based on a low-energy and
medium-baseline experiment. It is demonstrated that a relatively 
pure signal of leptonic CP violation at the percent level can be established
from such accelerator neutrino experiments.

{\Large\bf 2} ~ 
Let us denote the mass matrices of charged leptons and neutrinos in
vacuum to be $M_l$ and $M_\nu$, respectively. 
The phenomenon of lepton flavor mixing arises from the mismatch between
the diagonalization of $M_l$ and that of $M_\nu$ in an arbitrary flavor 
basis. Without loss of generality, one can choose to identify the flavor
eigenstates of charged leptons with their mass eigenstates. In this
specific basis, where $M_l$ takes the diagonal form
$\overline{M}_l \equiv {\rm Diag} \{m_e, m_\mu, m_\tau \}$,
the corresponding lepton flavor mixing matrix $V$  
links the neutrino mass eigenstates $(\nu_1, \nu_2, \nu_3)$ 
to the neutrino flavor eigenstates $(\nu_e, \nu_\mu, \nu_\tau)$.
The effective Hamiltonian responsible for the propagation of neutrinos
in vacuum can be written as \cite{Wolfenstein}
\begin{equation}
{\cal H}_{\rm eff} \; =\; \frac{1}{2E} 
\left (V \overline{M}^2_\nu V^\dagger \right ) \; ,
\end{equation}
where $\overline{M}_\nu \equiv {\rm Diag}\{m_1, m_2, m_3 \}$ with
$m_i$ being the neutrino mass eigenvalues, and $E \gg m_i$ denotes
the neutrino beam energy. 
When neutrinos travel through a normal material medium (e.g., the earth), 
which consists of electrons but of no muons or taus, they encounter
both charged- and neutral-current interactions with electrons. 
The neutral-current interactions are universal for $\nu_e$, $\nu_\mu$ 
and $\nu_\tau$ neutrinos, therefore they lead only to an overall 
and unobservable phase for neutrino mixing. The charged-current
interactions are likely to modify the features of neutrino mixing in vacuum
and must be taken into account in all realistic neutrino oscillation
experiments. Let us use ${\bf M}_\nu$ and $\bf V$ to denote 
the effective neutrino mass matrix and the effective flavor mixing matrix 
in matter. Then the effective Hamiltonian responsible for the propagation 
of neutrinos in matter can be written as \cite{Wolfenstein}
\begin{equation}
{\cal H}^{\rm m}_{\rm eff} \; =\; \frac{1}{2E} 
\left ({\bf V} \overline{\bf M}^2_\nu {\bf V}^\dagger \right ) \; ,
\end{equation}
in which 
$\overline{\bf M}_\nu \equiv {\rm Diag}\{{\bf m}_1, {\bf m}_2, {\bf m}_3 \}$ 
with ${\bf m}_i$ being the effective neutrino mass eigenvalues in matter.
The deviation of ${\cal H}^{\rm m}_{\rm eff}$ from ${\cal H}_{\rm eff}$,
denoted as $\Delta {\cal H}_{\rm eff}$, is given by
\begin{equation}
\Delta {\cal H}_{\rm eff} \; =\; \left ( \matrix{
{\bf A} & 0 & 0 \cr
0 & 0 & 0 \cr
0 & 0 & 0 \cr} \right ) \; ,
\end{equation}
in which ${\bf A} = \sqrt{2} ~ G_{\rm F} N_e$ measures the
charged-current contribution to the $\nu_e e^-$ forward scattering,
and $N_e$ is the background density of electrons.
Assuming a constant earth density profile 
(i.e., $N_e$ = constant), which is quite close to the reality
for most of the long- and medium-baseline 
neutrino experiments \cite{LB}, one can derive the analytical 
relationship between ${\bf m}_i$ and $m_i$ as well as that between 
$\bf V$ and $V$ from Eqs. (1) and (2) \cite{Xing00}. Instead of
repeating such calculations, we list the relevant results for 
${\bf m}_i$ and $\bf V$ in Appendix A. Subsequently we concentrate
on the commutators of lepton mass matrices and explore how leptonic CP 
violation in matter is related to that in vacuum. Our discussion in
this section is essentially the extension of that in Ref. \cite{Scott}. 

An instructive measure of the lepton flavor mixing, i.e., the mismatch 
between the diagonalization of $M_l$ and that of $M_\nu$ (or ${\bf M}_\nu$), 
is the commutators of lepton mass matrices, which can be defined as
\begin{eqnarray}
\left [ M_l M^\dagger_l ~ , ~ M_\nu M^\dagger_\nu \right ] 
& \equiv & i X \; ,
\nonumber \\
\left [ M_l M^\dagger_l ~ , ~ M^\dagger_\nu M_\nu \right ] 
& \equiv & i \overline{X} \; ,
\nonumber \\
\left [ M^\dagger_l M_l ~ , ~ M_\nu M^\dagger_\nu \right ] 
& \equiv & i Y \; ,
\nonumber \\
\left [ M^\dagger_l M_l ~ , ~ M^\dagger_\nu M_\nu \right ] 
& \equiv & i \overline{Y} \; 
\end{eqnarray}
for neutrinos propagating through vacuum; or as
\begin{eqnarray}
\left [ M_l M^\dagger_l ~ , ~ {\bf M}_\nu {\bf M}^\dagger_\nu \right ] 
& \equiv & i {\bf X} \; ,
\nonumber \\
\left [ M_l M^\dagger_l ~ , ~ {\bf M}^\dagger_\nu {\bf M}_\nu \right ] 
& \equiv & i \overline{\bf X} \; ,
\nonumber \\
\left [ M^\dagger_l M_l ~ , ~ {\bf M}_\nu {\bf M}^\dagger_\nu \right ] 
& \equiv & i {\bf Y} \; ,
\nonumber \\
\left [ M^\dagger_l M_l ~ , ~ {\bf M}^\dagger_\nu {\bf M}_\nu \right ] 
& \equiv & i \overline{\bf Y} \; 
\end{eqnarray} 
for neutrinos interacting with matter.
In the flavor basis where $M_l = \overline{M}_l$ holds,
we obtain $X = Y$, $\overline{X} = \overline{Y}$ and
${\bf X} = {\bf Y}$, $\overline{\bf X} = \overline{\bf Y}$.
It is important to note that 
\begin{eqnarray}
M_\nu M^\dagger_\nu & = & V \overline{M}^2_\nu V^\dagger 
\; = \; 2 E {\cal H}_{\rm eff} \; ,
\nonumber \\
{\bf M}_\nu {\bf M}^\dagger_\nu & = & {\bf V} \overline{\bf M}^2_\nu 
{\bf V}^\dagger \; = \; 2 E {\cal H}^{\rm m}_{\rm eff} 
\end{eqnarray}
hold in the chosen basis, no matter whether neutrinos are Dirac or
Majorana particles (see Appendix B for a brief proof). In contrast,
there is no direct relationship between $M^\dagger_\nu M_\nu$ (or
${\bf M}^\dagger_\nu {\bf M}_\nu$) and ${\cal H}_{\rm eff}$ (or
${\cal H}^{\rm m}_{\rm eff}$). We are therefore interested only in
the commutators $X$ and ${\bf X}$, which can be rewritten as 
\begin{eqnarray}
X & = & i \left [V \overline{M}^2_\nu V^\dagger ~ , 
~ \overline{M}^2_l \right ] \; , 
\nonumber \\
{\bf X} & = & i \left [{\bf V} \overline{\bf M}^2_\nu {\bf V}^\dagger ~ , 
~ \overline{M}^2_l \right ] \; .
\end{eqnarray}
The determinants of $X$ and ${\bf X}$ are related to leptonic CP violation 
in vacuum and in matter, respectively. To see this point clearly, we
carry out a straightforward calculation and arrive at
\begin{eqnarray}
{\rm Det}(X) & = & 2 J 
\prod_{\alpha < \beta} \left (m^2_\alpha - m^2_\beta \right )
\prod_{i<j} \left (m^2_i - m^2_j \right ) \; ,
\nonumber \\
{\rm Det}({\bf X}) & = & 2 {\bf J} 
\prod_{\alpha < \beta} \left (m^2_\alpha - m^2_\beta \right )
\prod_{i<j} \left ({\bf m}^2_i - {\bf m}^2_j \right ) \; ,
\end{eqnarray}
where the Greek indices run over $(e, \mu, \tau)$; the
Latin indices run over $(1, 2, 3)$; $J$ and $\bf J$ are the universal 
CP-violating parameters of $V$ and $\bf V$, respectively.
In vacuum $J$ is defined through \cite{Jarlskog} 
\begin{equation}
{\rm Im} \left (V_{\alpha i}V_{\beta j} V^*_{\alpha j}V^*_{\beta i} \right )
\; = \; J \sum_{\gamma,k} \epsilon^{~}_{\alpha \beta \gamma} 
\epsilon^{~}_{ijk} \; ,
\end{equation}
where $(\alpha, \beta, \gamma)$ and $(i, j, k)$ run over
$(e, \mu, \tau)$ and $(1, 2, 3)$, respectively. Similarly $\bf J$ can
be defined in terms of the matrix elements of $\bf V$.
We see that the quantities ${\rm Det}(X)$ and
${\rm Det}({\bf X})$ measure leptonic CP violation rephasing-invariantly.

We proceed to find out the relationship between the universal CP-violating
parameters $\bf J$ and $J$. In view of Eq. (6) as well as Eq. (3), 
we immediately realize
that 
\begin{equation}
{\bf X} \; = \; 2iE
\left [ {\cal H}^{\rm m}_{\rm eff} ~ , ~ \overline{M}^2_l \right ] 
\; = \; X 
+ 2iE \left [ \Delta {\cal H}_{\rm eff} ~ , ~ \overline{M}^2_l \right ]  
\; = \; X \; .
\end{equation}
This interesting result implies that the commutators
of lepton mass matrices are invariant under matter effects.
As a consequence, ${\rm Det}({\bf X}) = {\rm Det}(X)$, leading
to an elegant relationship between $\bf J$ and $J$ \cite{Scott}:
\begin{equation}
{\bf J} \prod_{i<j} \left ({\bf m}^2_i - {\bf m}^2_j \right ) \; = \;
J \prod_{i<j} \left (m^2_i - m^2_j \right ) \; .
\end{equation}
$\bf J$ depends on the matter effect (i.e., the parameter $\bf A$)
through ${\bf m}^2_i$. Of course ${\bf J} = J$ if ${\bf A} = 0$, and
${\bf J} = 0$ if $J =0$. 

The results obtained above are valid for neutrinos propagating in
vacuum and in matter. As for antineutrinos, the corresponding
formulas can straightforwardly be written out from Eqs. (1) -- (11)
through the replacements $V \Longrightarrow V^*$ and
${\bf A} \Longrightarrow (-{\bf A})$. 

{\Large\bf 3} ~
Now we illustrate the features of lepton flavor mixing and CP
violation in the scenario of low-energy 
(100 MeV $\leq E \leq$ 1 GeV) and medium-baseline 
(100 km $\leq L \leq$ 400 km) neutrino experiments.
As the large-angle MSW solution to the solar
neutrino problem seems to be favored by the latest Super-Kamiokande 
data \cite{Osaka}, we typically take the mass-squared difference 
$\Delta m^2_{\rm sun} = 5\cdot 10^{-5} ~ {\rm eV}^2$. We take
$\Delta m^2_{\rm atm} = 3\cdot 10^{-3} ~ {\rm eV}^2$ in view of the present
data on atmospheric neutrino oscillations. The corresponding 
pattern of lepton flavor mixing is expected to be nearly 
bi-maximal; 
i.e., $|V_{e3}| \ll 1$ as supported by the CHOOZ experiment \cite{CHOOZ}, 
$|V_{e1}|\sim |V_{e2}| \sim {\cal O}(1)$,
and $|V_{\mu 3}|\sim |V_{\tau 3}| \sim {\cal O}(1)$. 
The approximate decoupling of atmospheric and solar neutrino
oscillations implies that one can set
\begin{eqnarray}
\Delta m^2_{21} \equiv m^2_2 - m^2_1 
& \approx & \pm \Delta m^2_{\rm sun} \; ,
\nonumber \\
\Delta m^2_{31} \equiv m^2_3 - m^2_1 
& \approx & \pm \Delta m^2_{\rm atm} \; .
\end{eqnarray}
Of course $\Delta m^2_{32} \approx \Delta m^2_{31}$ holds in this 
approximation.
Note that the phenomena of flavor mixing and CP violation in neutrino 
oscillations can fully be described in terms of four independent 
parameters of $V$. To satisfy the nearly bi-maximal neutrino mixing with
large CP violation, we typically choose $|V_{e1}| = 0.816$, 
$|V_{e2}| = 0.571$, $|V_{\mu 3}| = 0.640$, and $J = \pm 0.020$ as the 
four independent input parameters in the numerical calculations
\footnote{This specific choice corresponds to $\theta_{12} \approx 35^\circ$,
$\theta_{23}\approx 40^\circ$, $\theta_{13}\approx 5^\circ$, and
$\delta \approx \pm 90^\circ$ in the standard parametrization of 
$V$ \cite{PDG}, in which $J = \sin\theta_{12} \cos\theta_{12} \sin\theta_{23}
\cos\theta_{23} \sin\theta_{13} \cos^2\theta_{13} \sin\delta$ holds.
$J>0$ and $J<0$ are associated with the case of neutrino mixing and 
that of antineutrino mixing, respectively.}.
In addition, the dependence of the terrestrial matter effect on the 
neutrino beam energy is given as 
$A \equiv 2E {\bf A}= 2.28\cdot 10^{-4} ~{\rm eV}^2 E$/[GeV] \cite{LB}.
With the help of Eq. (11) and those listed in Appendix A, we are then able 
to compute the ratios 
$\Delta {\bf m}^2_{i1}/\Delta m^2_{i1}$ (for $i=2,3$), 
$|{\bf V}_{\alpha i}|/|V_{\alpha i}|$ (for $\alpha = e, \mu, \tau$ and
$i = 1, 2, 3$), and ${\bf J}/J$ as functions of $E$. The relevant results
are shown respectively in Figs. 1, 2 and 3, in which $\Delta m^2_{21} >0$ and
$\Delta m^2_{31} > 0$ have been assumed. 
Some comments are in order.

(a) Fig.1 shows that the earth-induced matter effect on 
$\Delta m^2_{21}$ is significant, but that on $\Delta m^2_{31}$
is negligibly small in the chosen range of $E$ (i.e., 
$\Delta {\bf m}^2_{31} \approx \Delta m^2_{31}$ is an acceptable
approximation). In addition,
the mass-squared differences of neutrinos are relatively
more sensitive to the matter effect than those of antineutrinos.

(b) Fig. 2 shows that the matter effects on $|V_{\mu 3}|$
and $|V_{\tau 3}|$ are negligible in the low-energy neutrino
experiment. The smallest matrix element $|V_{e3}|$ is weakly
sensitive to the matter effect; e.g., $|{\bf V}_{e3}|/|V_{e3}|$
deviates about $7\%$ from unity for $E\approx 1$ GeV. In contrast,
the other six matrix elements of $V$ are significantly modified by the
terrestrial matter effects. For $|V_{e1}|$, $|V_{\mu 2}|$ and
$|V_{\tau 2}|$, the relevant matter effects of neutrinos are
more important than those of antineutrinos. For $|V_{e2}|$,
$|V_{\mu 1}|$ and $|V_{\tau 1}|$, the matter effects of neutrinos
and antineutrinos are essentially comparable in magnitude.

(c) Fig. 3 shows that the magnitude of $\bf J$ may rapidly decrease 
with the neutrino beam energy $E$. This feature implies that the
low-energy and medium-baseline neutrino experiments are likely to
provide a good chance for measurements of leptonic CP-violating
and T-violating asymmetries. Note also that $\bf J$ undergoes a
resonance around $E\sim 100$ MeV because of the terrestrial matter
effect. 

So far we have chosen $\Delta m^2_{21} >0$ and $\Delta m^2_{31} > 0$
in the numerical calculations. As the spectrum of neutrino masses
is unknown, it is possible that $\Delta m^2_{21} <0$ and (or)
$\Delta m^2_{31} <0$. After a careful analysis of the dependence
of $\bf J$ on positive and negative values of $\Delta m^2_{21}$
and $\Delta m^2_{31}$, we find the following exact relations:
\begin{eqnarray}
{\bf J} (+\Delta m^2_{21}, +\Delta m^2_{31}, +A )
& = & -{\bf J} (-\Delta m^2_{21}, -\Delta m^2_{31}, -A ) \; ,
\nonumber \\
{\bf J} (+\Delta m^2_{21}, -\Delta m^2_{31}, +A )
& = & -{\bf J} (-\Delta m^2_{21}, +\Delta m^2_{31}, -A ) \; ,
\nonumber \\
{\bf J} (+\Delta m^2_{21}, +\Delta m^2_{31}, -A )
& = & -{\bf J} (-\Delta m^2_{21}, -\Delta m^2_{31}, +A ) \; ,
\nonumber \\
{\bf J} (+\Delta m^2_{21}, -\Delta m^2_{31}, -A )
& = & -{\bf J} (-\Delta m^2_{21}, +\Delta m^2_{31}, +A ) \; .
\end{eqnarray}
The validity of these equations, which are independent of both
the neutrino beam energy and the baseline length, can easily be proved
with the help of Eqs. (11), (A1) and (A2). The key point
is that $\Delta {\bf m}^2_{21}$ and $\Delta {\bf m}^2_{31}$ keep 
unchanged in proper arrangements of the signs for $\Delta m^2_{21}$, 
$\Delta m^2_{31}$ and $A$. Furthermore, there exist four
approximate relations:
\begin{eqnarray}
{\bf J} (+\Delta m^2_{21}, +\Delta m^2_{31}, +A )
& \approx & {\bf J} (+\Delta m^2_{21}, -\Delta m^2_{31}, +A ) \; ,
\nonumber \\
{\bf J} (+\Delta m^2_{21}, +\Delta m^2_{31}, -A )
& \approx & {\bf J} (+\Delta m^2_{21}, -\Delta m^2_{31}, -A ) \; ,
\nonumber \\
{\bf J} (-\Delta m^2_{21}, +\Delta m^2_{31}, -A )
& \approx & {\bf J} (-\Delta m^2_{21}, -\Delta m^2_{31}, -A ) \; ,
\nonumber \\
{\bf J} (-\Delta m^2_{21}, -\Delta m^2_{31}, +A )
& \approx & {\bf J} (-\Delta m^2_{21}, +\Delta m^2_{31}, +A ) \; ,
\end{eqnarray}
which hold to a good degree of accuracy (with the relative errors 
$\leq 15\%$) for the chosen neutrino beam energy.
This result implies that changing the sign of $\Delta m^2_{31}$
does not affect the magnitude of $\bf J$ significantly
\footnote{Note that the sign of 
$\Delta m^2_{32}$ (= $\Delta m^2_{31} - \Delta m^2_{21}$) changes 
simultaneously with that of $\Delta m^2_{31}$, simply because  
$|\Delta m^2_{32}| \approx |\Delta m^2_{31}| \gg |\Delta m^2_{21}|$ holds.
The sensitivity of ${\bf J}/J$ to the signs of $\Delta m^2_{21}$
and $\Delta m^2_{31}$ has been examined in Ref. \cite{Yokomakura}.}.
We expect that those interesting relations in Eqs. (13) and (14) can 
experimentally be tested in the near future.

In a similar way we have carefully analyzed the behaviors of
$|{\bf V}_{\alpha i}|$ (for $\alpha = e, \mu, \tau$ and
$i =1,2,3$) with respect to the negative values of $\Delta m^2_{21}$ and
$\Delta m^2_{31}$. Instead of presenting the details
of our quantitative results, we only make two qualitative remarks:
(1) while $\bf J$ is not sensitive to the sign
of $\Delta m^2_{31}$, $|{\bf V}_{\alpha i}|$ may
dramatically be suppressed or enhanced by changing
the sign of $\Delta m^2_{31}$; (2) in contrast,
the dependence of $|{\bf V}_{\alpha i}|$ on the
sign of $\Delta m^2_{21}$ is less significant. 

Finally it is worth mentioning that dramatic changes of the 
results shown in Figs. 1 -- 3 do not happen, even if one allows
every parameter to change in a reasonable region around the 
originally chosen value (e.g., 
$\Delta m^2_{21} \sim (10^{-5} - 10^{-4}) ~ {\rm eV}^2$,
$\Delta m^2_{31} \sim (1 - 6)\cdot 10^{-3} ~ {\rm eV}^2$,
$|V_{e1}| \sim 0.7 - 0.9$,
$|V_{e2}| \sim 0.5 - 0.7$,
$|V_{\mu 3}| \sim 0.5 - 0.8$,
and $|J| \sim 0.01 - 0.03$).
This observation implies that the matter-corrected quantities
of lepton flavor mixing and CP violation have quite stable
behaviors, as a qualitative consequence of the nearly bi-maximal
mixing pattern of $V$ and the rough hierarchy 
$\Delta m^2_{\rm sun} < A < \Delta m^2_{\rm atm}$ in the 
experimental scenario under consideration.

{\Large\bf 4} ~
Let us turn to CP- and T-violating asymmetries in the low-energy and
medium-baseline neutrino oscillation experiments. 
The conversion probability of a neutrino $\nu_\alpha$
to another neutrino $\nu^{~}_\beta$ in vacuum is given by \cite{Xing00}
\begin{equation}
P(\nu_\alpha \rightarrow \nu^{~}_\beta) \; = \;
-4 \sum_{i<j} [ {\rm Re} ( V_{\alpha i} V_{\beta j} 
V^*_{\alpha j} V^*_{\beta i} ) \cdot \sin^2 F_{ij} ] 
~ + ~ 8 J \prod_{i<j} \sin F_{ij} \; ,
\end{equation}
where $(\alpha, \beta)$ run over $(e, \mu)$, $(\mu, \tau)$ or $(\tau, e)$,
and $F_{ij} \equiv 1.27 \Delta m^2_{ij} L/E$ with $L$ being the
baseline length (in unit of km) and $E$ being the neutrino beam energy 
(in unit of GeV). 
The probabilities of $\nu^{~}_\beta \rightarrow \nu_\alpha$ and
$\overline{\nu}_\alpha \rightarrow \overline{\nu}^{~}_\beta$ transitions
can straightforwardly be read off from Eq. (15) with the replacement 
$J \Longrightarrow -J$. Clearly 
$P(\overline{\nu}_\alpha \rightarrow \overline{\nu}^{~}_\beta) =
P(\nu^{~}_\beta \rightarrow \nu_\alpha)$ 
holds as a straightforward consequence
of CPT invariance. The CP-violating asymmetry between
$P(\nu_\alpha \rightarrow \nu^{~}_\beta)$ and
$P(\overline{\nu}_\alpha \rightarrow \overline{\nu}^{~}_\beta)$
amounts to the T-violating asymmetry between
$P(\nu_\alpha \rightarrow \nu^{~}_\beta)$ and
$P(\nu^{~}_\beta \rightarrow \nu_\alpha)$ \cite{Cabibbo,FX00}:
\begin{eqnarray}
\Delta P & = & P(\nu_\alpha \rightarrow \nu^{~}_\beta) ~ - ~
P(\overline{\nu}_\alpha \rightarrow \overline{\nu}^{~}_\beta) \; 
\nonumber \\
& = & P(\nu_\alpha \rightarrow \nu^{~}_\beta) ~ - ~ 
P(\nu^{~}_\beta \rightarrow \nu_\alpha) \;
\nonumber \\
& = & -16 J \sin F_{21} \cdot \sin F_{31} \cdot \sin F_{32} \; .
\end{eqnarray}
Because of $|\Delta m^2_{31}| \approx |\Delta m^2_{32}| \gg |\Delta m^2_{21}|$,
a favorable signal of CP or T violation can be obtained only when
the condition $|\Delta m^2_{21}| \sim E/L$ is satisfied (i.e.,
$|\sin F_{21}| \sim {\cal O}(1)$ is acquired). Therefore only the large-angle
MSW solution to the solar neutrino problem is possible to meet such
a prerequisite for the observation of CP violation in realistic
neutrino oscillation experiments.

When the terrestrial matter effects are taken into account, 
the probability of the $\nu_\alpha \rightarrow \nu^{~}_\beta$ transition
becomes
\begin{equation}
{\bf P}(\nu_\alpha \rightarrow \nu^{~}_\beta) \; = \;
-4 \sum_{i<j} [ {\rm Re} ( {\bf V}_{\alpha i} {\bf V}_{\beta j} 
{\bf V}^*_{\alpha j} {\bf V}^*_{\beta i} ) \cdot \sin^2 {\bf F}_{ij} ] 
~ + ~ 8 {\bf J} \prod_{i<j} \sin {\bf F}_{ij} \; ,
\end{equation}
in which ${\bf F}_{ij} \equiv 1.27 \Delta {\bf m}^2_{ij} L/E$.
The probability ${\bf P}(\nu^{~}_\beta \rightarrow \nu_\alpha)$ 
can directly be read off from Eq. (17) with the replacement 
$J \Longrightarrow -J$. To obtain the probability
${\bf P}(\overline{\nu}_\alpha \rightarrow \overline{\nu}^{~}_\beta)$,
however, both the replacements $J \Longrightarrow -J$
and $A \Longrightarrow -A$ need be made for Eq. (17). 
In this case, 
${\bf P}(\overline{\nu}_\alpha \rightarrow \overline{\nu}^{~}_\beta)$
is not equal to ${\bf P}(\nu^{~}_\beta \rightarrow \nu_\alpha)$.
Such a false signal of CPT violation measures the matter effect!
Similar to Eq. (16), the CP- and T-violating asymmetries in the
presence of matter effects can be defined respectively as
\begin{eqnarray}
\Delta {\bf P} & = & {\bf P}(\nu_\alpha \rightarrow \nu^{~}_\beta) ~ - ~
{\bf P}(\overline{\nu}_\alpha \rightarrow \overline{\nu}^{~}_\beta) \; , 
\nonumber \\
\Delta \tilde{\bf P} & = & {\bf P}(\nu_\alpha \rightarrow \nu^{~}_\beta) ~ - ~ 
{\bf P}(\nu^{~}_\beta \rightarrow \nu_\alpha) \; ,
\end{eqnarray}
where the flavor indices $(\alpha, \beta)$ run over $(e, \mu)$,
$(\mu, \tau)$ or $(\tau, e)$.
In general, $\Delta \tilde{\bf P} \neq \Delta {\bf P}$ because of the
matter-induced corrections to $\Delta P$. 
Note that the overall matter effects residing 
in $\Delta \tilde{\bf P}$ is expected to be tiny. Indeed 
$\Delta \tilde{\bf P} \approx \Delta P$ has numerically been confirmed
to be an excellent approximation \cite{LB,FX00}. 
The reason is simply that the 
matter-induced corrections to $P(\nu_\alpha \rightarrow \nu^{~}_\beta)$
and $P(\nu^{~}_\beta \rightarrow \nu_\alpha)$, which depend on the same
parameter $\bf A$, may essentially cancel each other in the asymmetry
$\Delta \tilde{\bf P}$. In contrast, 
${\bf P}(\nu_\alpha \rightarrow \nu^{~}_\beta)$ and
${\bf P}(\overline{\nu}_\alpha \rightarrow \overline{\nu}^{~}_\beta)$
are associated separately with $(+{\bf A})$ and $(-{\bf A})$, thus
there is no large cancellation of matter effects in the asymmetry
$\Delta {\bf P}$. 

To illustrate, we perform a numerical analysis of the CP-violating 
asymmetry $\Delta {\bf P}$ between $\nu_e \rightarrow \nu_\mu$
and $\overline{\nu}_e \rightarrow \overline{\nu}_\mu$ transitions
as well as the T-violating asymmetry $\Delta \tilde{\bf P}$ 
between $\nu_e \rightarrow \nu_\mu$ and $\nu_\mu \rightarrow \nu_e$ 
transitions. The values of the relevant input parameters are
taken the same as before. We first choose $\Delta m^2_{21}>0$ and
$\Delta m^2_{31}>0$ to calculate $\Delta P$ and $\Delta {\bf P}$ as
functions of the beam energy $E$ with respect to 
$L=100$ km, 200 km, 300 km and 400 km. The results are shown in
Fig. 4. For the chosen range of $E$, one can see that the matter effect
on $\Delta P$ is insignificant; i.e., $\Delta {\bf P} \approx \Delta P$
is a good approximation. The largest deviation of $\Delta {\bf P}$ from
$\Delta P$, of the magnitude $0.12\%$ or so, takes place when $E$ and
$L$ satisfy the rough condition $L/E \sim 500$ km/GeV. This observation
is of course dependent on the values of the input parameters. We confirm
that $\Delta \tilde{\bf P} \approx \Delta P$ holds to a better degree
of accuracy than $\Delta {\bf P} \approx \Delta P$. 
Following a similar analysis, we find that 
the CP-violating asymmetry between $\nu_\mu \rightarrow \nu_e$ and 
$\overline{\nu}_\mu \rightarrow \overline{\nu}_e$ transitions amounts 
essentially to ($-\Delta {\bf P}$). It becomes clear that
a relatively pure CP-violating asymmetry at the percent level
can be obtained from such a low-energy and medium-baseline neutrino
experiment. 

The continuous dependence of the CP-violating asymmetry 
$\Delta {\bf P}$ on the baseline length $L$ is shown in Fig. 5
for five different values of the neutrino beam energy $E$ (i.e.,
$E= 100$ MeV, 200 MeV, 300 MeV, 400 MeV and 500 MeV). 
Fixing $E = 100$ MeV, for example, we observe that the maximal magnitude of 
$\Delta {\bf P}$ may reach $4\%$ at $L\sim 200$ km and
$6\%$ at $L\sim 280$ km. Note that $L$ should not be too large in
realistic experiments, in order to keep the neutrino beam opening angles
as small as possible \cite{Minakata}. 
For a specific experiment, one can select
the optimum values of $E$ and $L$ after taking its luminosity and 
other technical details into account. 

Finally let us examine the sensitivity of $\Delta {\bf P}$ to the
signs of $\Delta m^2_{21}$ and $\Delta m^2_{31}$. To be explicit,
we take $L=100$ km and 400 km. The dependence of $\Delta {\bf P}$ 
on $E$ is then studied with respect to four possible combinations 
for the signs of $\Delta m^2_{21}$ and $\Delta m^2_{31}$. The 
numerical results are depicted in Fig. 6. We find that the
following relations hold to a good degree of accuracy:
\begin{eqnarray}
\Delta {\bf P} (+\Delta m^2_{21}, + \Delta m^2_{31}) & \approx &
-\Delta {\bf P} (-\Delta m^2_{21}, - \Delta m^2_{31}) \; ,
\nonumber \\
\Delta {\bf P} (+\Delta m^2_{21}, - \Delta m^2_{31}) & \approx &
-\Delta {\bf P} (-\Delta m^2_{21}, + \Delta m^2_{31}) \; .
\end{eqnarray}
To understand the above equation, we notice from
Eq. (16) that $\Delta P(+\Delta m^2_{21}, \pm \Delta m^2_{31})$
and $\Delta P(-\Delta m^2_{21}, \mp \Delta m^2_{31})$ have the
same magnitude but the opposite signs. A slight deviation of
$\Delta {\bf P}(\pm \Delta m^2_{21}, \pm \Delta m^2_{31})$ from
$\Delta P(\pm \Delta m^2_{21}, \pm \Delta m^2_{31})$ arises from
the terrestrial matter effects, which are very samll in the
experimental scenario under consideration. Therefore we arrive 
at the approximate relations in Eq. (19). Note also that the
sensitivity of $\Delta {\bf P}$ to the sign of $\Delta m^2_{31}$
is quite weak, as shown in Fig. 6. This result is certainly consistent
with that for ${\bf J}$ in Eq. (14); i.e., changing the sign of
$\Delta m^2_{31}$ does not affect $\bf J$ and $\Delta {\bf P}$
significantly.

{\Large\bf 5} ~
In this paper we have introduced the commutators
of lepton mass matrices to describe the phenomenon of lepton 
flavor mixing. Their relations to the effective Hamiltonian 
responsible for the propagation of neutrinos are independent of the
nature of neutrinos (Dirac or Majorana). 
The determinants of those commutators are 
invariant under matter effects, leading to an elegant relationship
between the universal CP-violating parameters in matter and 
in vacuum. 

We have illustrated the features of lepton flavor mixing and CP 
violation in the scenario of low-energy (100 MeV $\leq E \leq$ 1 GeV)
and medium-baseline (100 km $\leq L\leq$ 400 km) neutrino experiments,
In particular, the terrestrial matter effects on the elements of
the lepton mixing matrix and on the rephasing-invariant measure of CP 
violation are systematically analyzed. Some useful relations have been found 
for the matter-corrected parameters of CP or T violation in respect to
different signs of the neutrino mass-squared differences.

We have also presented a detailed analysis of CP- and T-violating 
asymmetries in neutrino oscillations, based on a medium-baseline
experiment with low-energy neutrino beams. The terrestrial matter
effects are demonstrated to be insignificant and sometimes even negligible 
in such an experimental scenario. A relatively pure signal of leptonic
CP violation at the percent level can be established from the
probability asymmetry between $\nu_\mu \rightarrow \nu_e$ and
$\overline{\nu}_\mu \rightarrow \overline{\nu}_e$ transitions, or
between $\nu_e \rightarrow \nu_\mu$ and
$\overline{\nu}_e \rightarrow \overline{\nu}_\mu$ transitions.

To realize a low-energy and medium-baseline neutrino experiment needs
high-intensity and narrow-band neutrino beams (e.g., 10 to 100 times
more intense than the neutrino flux in the K2K 
experiment \cite{Minakata}). 
Considering conventional neutrino beams produced from the charged
pion decay, for example, one may maximize the neutrino flux in the
forward direction by restricting the pion beam divergence.
The required radial focusing can be provided by a quadrupole channel
and (or) magnetic horns \cite{Richter,Barger}. The peak pion energy
and energy spread within the pion decay channel are determined by
the beamline optics, which in turn determine the neutrino spectrum.
If the optics are designed to accept a large pion momentum spread,
the resultant wide-band beam will contain a large neutrino flux
with a broad energy spectrum; and if the optics are designed to accept
a smaller pion momentum spread, the resultant narrow-band beam will
contain a smaller neutrino flux with a narrower energy spectrum \cite{Barger}.
A detailed study of such technical problems is certainly desirable, but 
beyond the scope of this work. 

We expect that the low-energy, medium-baseline neutrino
experiments and the
high-energy, long-baseline neutrino experiments may be
complementary to each other, in the near future, towards a precise 
determination of the flavor mixing and CP-violating parameters in
the lepton sector.

The author is indebted to A. Romanino for her enlightening comments on this
paper.

\newpage

\centerline{\Large APPENDIX A}
\vspace{0.5cm}

\setcounter{equation}{0}
\renewcommand{\thesection}{\Alph{section}}
\renewcommand{\theequation}{\thesection\arabic{equation}}
\setcounter{section}{1}

In the assumption of a constant earth density profile and
with the help of the effective Hamiltonians given in
in Eqs. (1) and (2), the author has recently derived
the exact and parametrization-independent formulas for
the matter-corrected neutrino masses ${\bf m}_i$ and the flavor
mixing matrix elements ${\bf V}_{\alpha i}$ in Ref. \cite{Xing00}.
The main results are briefly summarized as follows.

(1) The neutrino mass eigenvalues $m^{~}_i$ in vacuum and 
${\bf m}^{~}_i$ in matter are related to each other through 
\begin{eqnarray}
{\bf m}^2_1 & = & m^2_1 + \frac{1}{3} x - \frac{1}{3} \sqrt{x^2 - 3y}
\left [z + \sqrt{3 \left (1-z^2 \right )} \right ] , 
\nonumber \\ 
{\bf m}^2_2 & = & m^2_1 + \frac{1}{3} x - \frac{1}{3} \sqrt{x^2 -3y}
\left [z - \sqrt{3 \left (1-z^2 \right )} \right ] , 
\nonumber \\ 
{\bf m}^2_3 & = & m^2_1 + \frac{1}{3} x + \frac{2}{3} z \sqrt{x^2 - 3y} \; ,
\end{eqnarray}
where $x$, $y$ and $z$ are given by \cite{Zaglauer} 
\begin{eqnarray}
x & = & \Delta m^2_{21} + \Delta m^2_{31} + A \; , 
\nonumber \\
y & = & \Delta m^2_{21} \Delta m^2_{31} + A \left [ 
\Delta m^2_{21} \left ( 1 - |V_{e2}|^2 \right ) 
+ \Delta m^2_{31} \left ( 1 - |V_{e3}|^2 \right ) \right ] , 
\nonumber \\
z & = & \cos \left [ \frac{1}{3} \arccos \frac{2x^3 -9xy + 27
A \Delta m^2_{21} \Delta m^2_{31} |V_{e1}|^2}
{2 \left (x^2 - 3y \right )^{3/2}} \right ] 
\end{eqnarray}
with $\Delta m^2_{21}$ and $\Delta m^2_{31}$ defined in Eq. (12),
and $A \equiv 2E {\bf A} = 2\sqrt{2} ~ G_{\rm F} N_e E$. 
Of course, ${\bf m}^2_i = m^2_i$ can be reproduced from
Eq. (A1) if $A =0$ is taken. Only the mass-squared
differences $\Delta {\bf m}^2_{21} \equiv {\bf m}^2_2 - {\bf m}^2_1$ 
and $\Delta {\bf m}^2_{31} \equiv {\bf m}^2_3 - {\bf m}^2_1$ are
relevant to the practical neutrino oscillations in matter.

At this point it is worth mentioning that $m^2_i$, $|V_{ei}|^2$,
$A$ and ${\bf m}^2_i$ are correlated with one another via an
interesting equation, which has not been noticed in the literature:
\begin{equation}
A \left (m^2_i - m^2_k \right ) \left (m^2_j - m^2_k \right )
|V_{ek}|^2 \; =\; \prod^3_{n=1} \left ({\bf m}^2_n - m^2_k \right ) \; ,
\end{equation}
where $(i, j, k)$ run over $(1, 2, 3)$ with $i\neq j \neq k$. 
Assuming $|\Delta m^2_{31}| \gg |\Delta m^2_{21}|$ and taking
$k=3$, one can easily reproduce the
approximate analytical result for the correlation between
${\bf m}^2_i$ and $m^2_i$ obtained in Ref. \cite{Pantaleone}.

(2) The analytical relationship between the elements of $\bf V$ in
matter and those of $V$ in vacuum reads:
\begin{eqnarray}
{\bf V}_{\alpha i} \; = \; \frac{N_i}{D_i} V_{\alpha i} 
~ + ~ \frac{A}{D_i} V_{e i} \left [ \left ({\bf m}^2_i - m^2_j \right )
V^*_{e k} V_{\alpha k} + \left ({\bf m}^2_i - m^2_k \right )
V^*_{e j} V_{\alpha j} \right ] ,
\end{eqnarray}
in which $\alpha$ runs over $(e, \mu, \tau)$ and $(i, j, k)$ over $(1, 2, 3)$
with $i \neq j \neq k$, and
\begin{eqnarray}
N_i & = & \left ({\bf m}^2_i - m^2_j \right ) \left ({\bf m}^2_i - m^2_k \right )
- A \left [\left ({\bf m}^2_i - m^2_j \right ) |V_{e k}|^2  
+ \left ({\bf m}^2_i - m^2_k \right ) |V_{e j}|^2 \right ] ,
\nonumber \\
D^2_i & = & N^2_i + A^2 |V_{e i}|^2 \left [ 
\left ({\bf m}^2_i - m^2_j \right )^2 |V_{e k}|^2  
+ \left ({\bf m}^2_i - m^2_k \right )^2 |V_{e j}|^2 \right ] . 
\end{eqnarray}
Obviously $A=0$ leads to 
${\bf V}_{\alpha i} = V_{\alpha i}$. This exact and compact formula shows 
clearly how the flavor mixing matrix in vacuum is corrected by the matter 
effects. Instructive analytical approximations can be made for Eq. (A4),
once the spectrum of neutrino masses is experimentally known or
theoretically predicted.

The results listed above are valid for neutrinos interacting with matter.
As for antineutrinos propagating through matter, 
the relevant formulas can straightforwardly be
obtained from Eqs. (A1) -- (A5) through the replacements
$V\Longrightarrow V^*$ and $A\Longrightarrow -A$.

\vspace{2cm}
\centerline{\Large APPENDIX B}
\vspace{0.5cm}

\setcounter{equation}{0}
\renewcommand{\thesection}{\Alph{section}}
\renewcommand{\theequation}{\thesection\arabic{equation}}
\setcounter{section}{2}

This Appendix aims to show why 
$M_\nu M^\dagger_\nu = V \overline{M}^2_\nu V^\dagger$ holds in the flavor
basis where the charged lepton mass matrix is diagonal 
(i.e., $M_l = \overline{M}_l$), no matter whether neutrinos are Dirac
or Majorana particles. For simplicity, we focus on the standard 
charged-current weak interactions, in which only the left-handed 
leptons take part:
\begin{equation}
-{\cal L}_{\rm weak} \; =\; \frac{g}{\sqrt{2}} ~ 
\overline{(\nu_e, \nu_\mu, \nu_\tau)^{~}_{\rm L}} ~ \gamma^\mu
\left (\matrix{
e \cr
\mu \cr
\tau \cr} \right )_{\rm L} W^+_\mu ~ + ~ {\rm h.c.} \; ,
\end{equation}
where the flavor eigenstates of charged leptons are identified with
their mass eigenstates. As the lepton flavor mixing matrix $V$
is defined to link the neutrino flavor eigenstates 
$(\nu_e, \nu_\mu, \nu_\tau)$ to the neutrino mass eigenstates 
$(\nu_1, \nu_2, \nu_3)$, we get
\begin{equation}
-{\cal L}_{\rm weak} \; =\; \frac{g}{\sqrt{2}} ~ 
\overline{(\nu^{~}_1, \nu^{~}_2, \nu^{~}_3)^{~}_{\rm L}} ~ V^\dagger \gamma^\mu
\left (\matrix{
e \cr
\mu \cr
\tau \cr} \right )_{\rm L} W^+_\mu ~ + ~ {\rm h.c.} \; 
\end{equation}
in the chosen flavor basis. 

If neutrinos are Dirac particles, the mass term can be written as
\begin{equation}
- {\cal L}_{\rm Dirac} \; =\; 
\overline{(\nu_e, \nu_\mu, \nu_\tau)^{~}_{\rm L}} 
~ M_\nu \left (\matrix{
\nu_e \cr
\nu_\mu \cr
\nu_\tau \cr} \right )_{\rm R} + ~ {\rm h.c.} \; , 
\end{equation}
where $M_\nu$ is in general an arbitrary $3\times 3$ matrix. It is
always possible to diagonalize $M_\nu$ by a bi-unitary transformation:
$V^\dagger M_\nu \tilde{V} = \overline{M}_\nu$, where $V$ is just
the flavor mixing matrix of Dirac neutrinos consistent with Eq. (B2).
Obviously $M_\nu M^\dagger_\nu = V \overline{M}^2_\nu V^\dagger$ holds.

If neutrinos are Majorana particles, the mass terms turns out to be
\begin{equation}
-{\cal L}_{\rm Majorana} \; =\; \frac{1}{2} ~
\overline{(\nu_e, \nu_\mu, \nu_\tau)^{~}_{\rm L}}
~ M_\nu \left (\matrix{
\nu^{\rm c}_e \cr
\nu^{\rm c}_\mu \cr
\nu^{\rm c}_\tau \cr} \right )_{\rm R} + ~ {\rm h.c.} \; ,
\end{equation}
in which $\nu^{\rm c} \equiv C \overline{\nu}^{\rm T}$ with $C$ 
being the charge-conjugation operator. It is well known that $M_\nu$
must be a symmetric matrix and can be diagonalized by a single 
unitary transformation: $V^\dagger M_\nu V^* = \overline{M}_\nu$,
where $V$ is just the flavor mixing matrix of Majorana neutrinos 
consistent with Eq. (B2)
\footnote{If the Majorana mass matrix $M_\nu$ is diagonalized by the
transformation $U^{\rm T} M_\nu U = \overline{M}_\nu$, one will see
that it is $U^*$ (instead of $U$) linking the flavor eigenstates
to the mass eigenstates of neutrinos. Therefore the flavor mixing
matrix turns out to be $V = U^*$ in the chosen flavor basis.}.
Once again we arrive at
$M_\nu M^\dagger_\nu = V \overline{M}^2_\nu V^\dagger$.

A more general neutrino mass Lagrangian involves both Dirac and
Majorana terms \cite{Bilenky}. In this case, one can similarly
prove that the mass matrix of light (active) Majorana neutrinos
is symmetric and satisfies 
$M_\nu M^\dagger_\nu = V \overline{M}^2_\nu V^\dagger$. Thus 
${\cal H}_{\rm eff} = (M_\nu M^\dagger_\nu)/(2E)$ holds in vacuum, 
no matter whether neutrinos are Dirac or Majorana particles. 
In matter we have an analogous relation between 
${\cal H}^{\rm m}_{\rm eff}$ and ${\bf M}_\nu {\bf M}^\dagger_\nu$, 
as given in Eq. (6).

\newpage 

\begin{figure}[t]
\vspace{1.8cm}
\epsfig{file=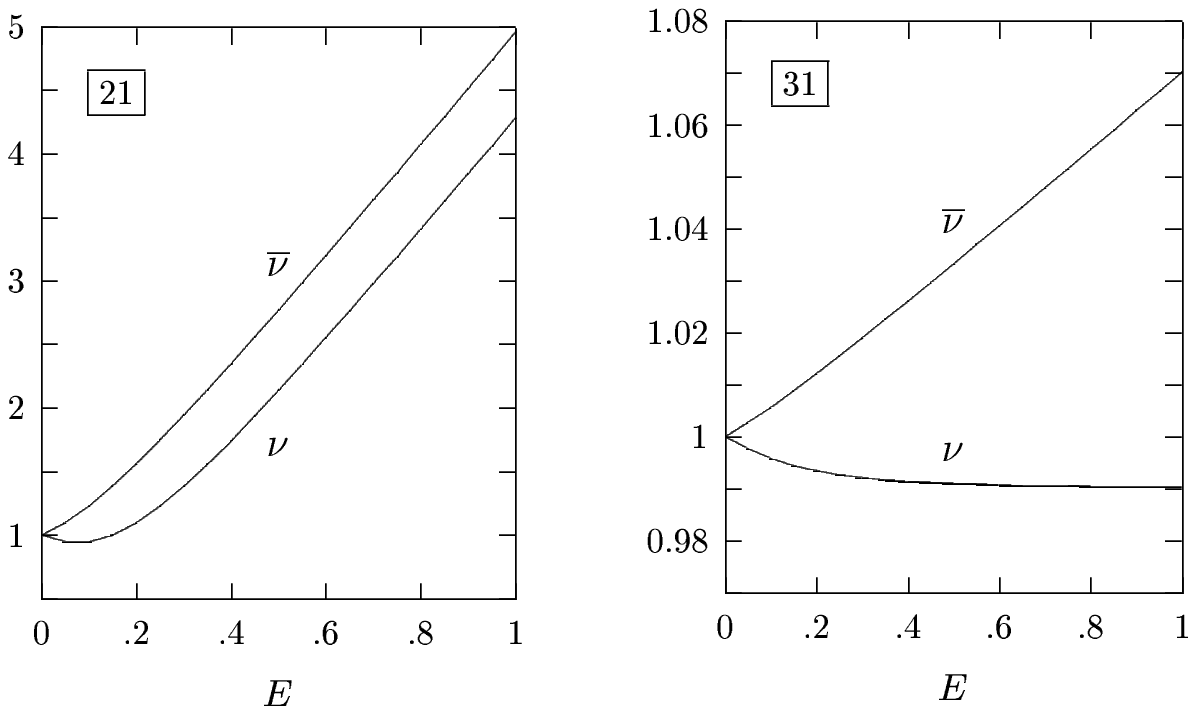,bbllx=2cm,bblly=6cm,bburx=19cm,bbury=28cm,%
width=15cm,height=20cm,angle=0,clip=}
\vspace{-12.5cm}
\caption{\small Ratios $\Delta {\bf m}^2_{21}/\Delta m^2_{21}$
and $\Delta {\bf m}^2_{31}/\Delta m^2_{31}$ changing with the 
beam energy $E$ (in unit of GeV) for neutrinos ($\nu$) and 
antineutrinos ($\overline{\nu}$),
in which $\Delta m^2_{21} = 5\cdot 10^{-5} ~ {\rm eV}^2$,
$\Delta m^2_{31} = 3\cdot 10^{-3} ~ {\rm eV}^2$,
$|V_{e1}| = 0.816$, and $|V_{e2}| = 0.571$ 
have typically been input.}
\end{figure}

\begin{figure}[t]
\vspace{2.5cm}
\epsfig{file=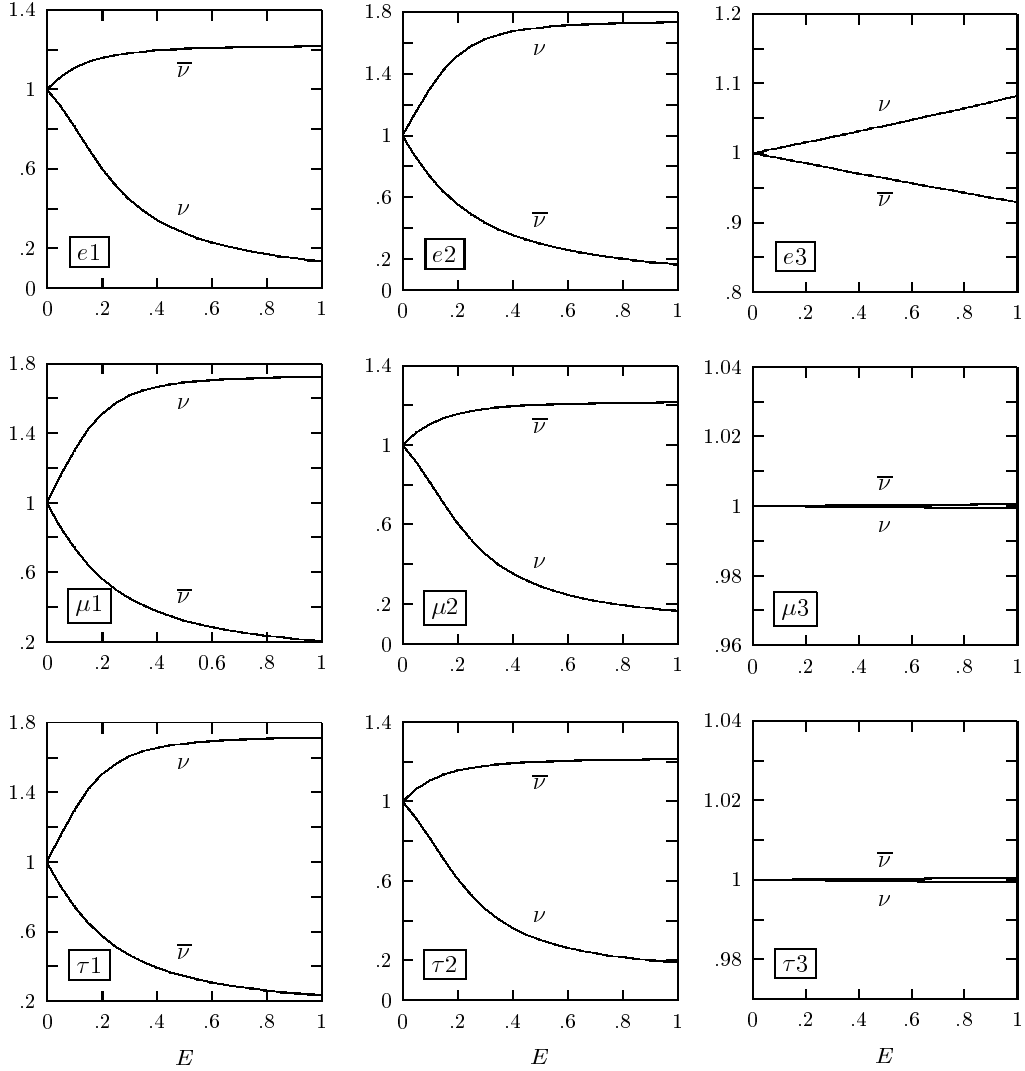,bbllx=1cm,bblly=5.8cm,bburx=19cm,bbury=30.cm,%
width=15cm,height=19cm,angle=0,clip=}
\vspace{-2.0cm}
\caption{\small Ratios $|{\bf V}_{\alpha i}|/|V_{\alpha i}|$ 
(for $\alpha = e, \mu, \tau$ and $i=1,2,3$)
changing with the beam energy $E$ (in unit of GeV) for neutrinos ($\nu$) 
and antineutrinos ($\overline{\nu}$),
in which $\Delta m^2_{21} = 5\cdot 10^{-5} ~ {\rm eV}^2$,
$\Delta m^2_{31} = 3\cdot 10^{-3} ~ {\rm eV}^2$,
$|V_{e1}| = 0.816$, $|V_{e2}| = 0.571$,
$|V_{\mu 3}| = 0.640$, and $J = \pm 0.020$  
have typically been input.}
\end{figure}

\begin{figure}[t]
\vspace{2.5cm}
\epsfig{file=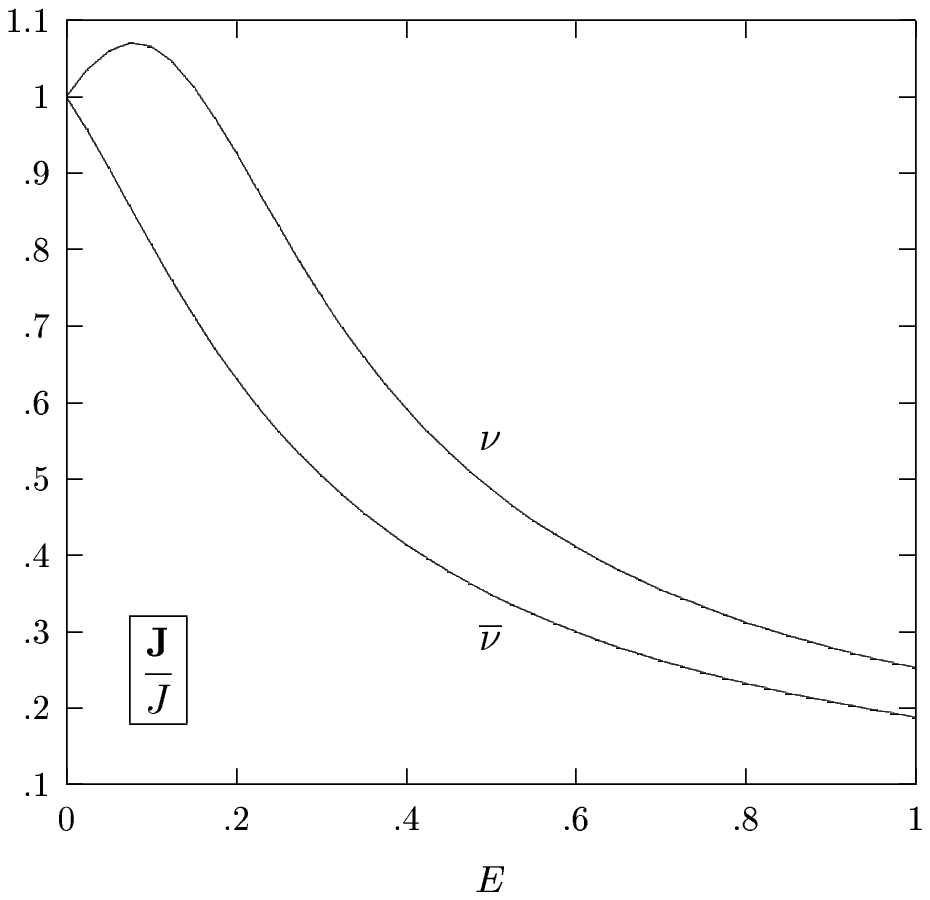,bbllx=2cm,bblly=6cm,bburx=19cm,bbury=28cm,%
width=15cm,height=20cm,angle=0,clip=}
\vspace{-10.45cm}
\caption{\small The ratio ${\bf J}/J$ 
changing with the beam energy $E$ (in unit of GeV) for neutrinos ($\nu$) and 
antineutrinos ($\overline{\nu}$),
in which $\Delta m^2_{21} = 5\cdot 10^{-5} ~ {\rm eV}^2$,
$\Delta m^2_{31} = 3\cdot 10^{-3} ~ {\rm eV}^2$,
$|V_{e1}| = 0.816$, $|V_{e2}| = 0.571$,
$|V_{\mu 3}| = 0.640$, and $J = \pm 0.020$  
have typically been input.}
\end{figure}

\begin{figure}[t]
\vspace{-0.8cm}
\epsfig{file=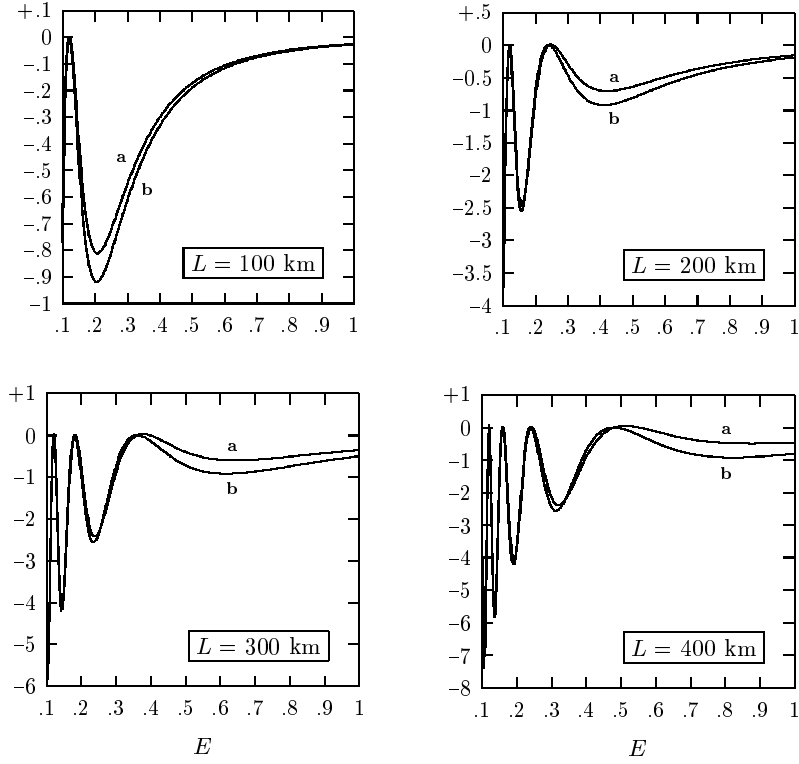,bbllx=1cm,bblly=5.8cm,bburx=19cm,bbury=30.cm,%
width=15cm,height=20cm,angle=0,clip=}
\vspace{-7.05cm}
\caption{\small The CP-violating
asymmetries between $\nu_e \rightarrow \nu_\mu$
and $\overline{\nu}_e \rightarrow \overline{\nu}_\mu$ transitions
in matter ({\bf a}: $\Delta {\bf P}$
in unit of $10^{-2}$) and in vacuum ({\bf b}: $\Delta P$ in unit of
$10^{-2}$) changing with the beam energy $E$ (in unit of GeV),
in which $\Delta m^2_{21} = 5\cdot 10^{-5} ~ {\rm eV}^2$,
$\Delta m^2_{31} = 3\cdot 10^{-3} ~ {\rm eV}^2$,
$|V_{e1}| = 0.816$, $|V_{e2}| = 0.571$,
$|V_{\mu 3}| = 0.640$, and $J = \pm 0.020$  
have typically been input.}
\end{figure}

\begin{figure}[t]
\vspace{-0.8cm}
\epsfig{file=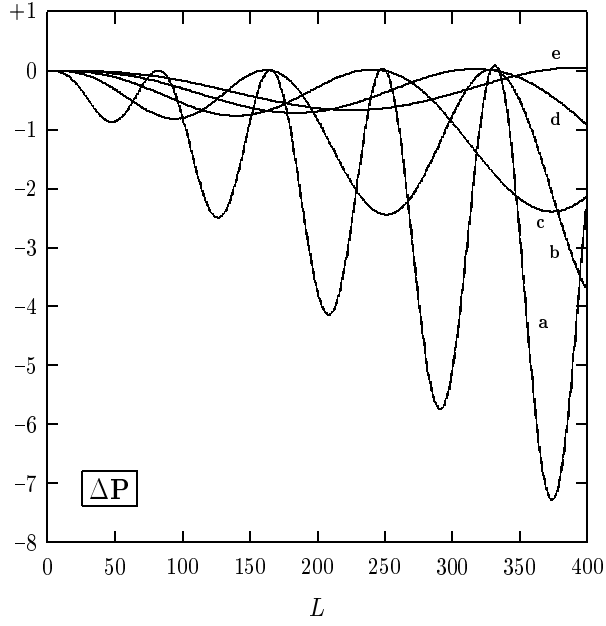,bbllx=0cm,bblly=6cm,bburx=19cm,bbury=28cm,%
width=15.5cm,height=20cm,angle=0,clip=}
\vspace{-10.4cm}
\caption{\small The CP-violating asymmetry $\Delta {\bf P}$ between 
$\nu_e \rightarrow \nu_\mu$
and $\overline{\nu}_e \rightarrow \overline{\nu}_\mu$ transitions
(in unit of $10^{-2}$)
changing with the baseline length $L$ (in unit of km) and the
beam energy $E$ ({\bf a}: 100 MeV; {\bf b}: 200 MeV; {\bf c}: 300 MeV;
{\bf d}: 400 MeV; {\bf e}: 500 MeV). Here
$\Delta m^2_{21} = 5\cdot 10^{-5} ~ {\rm eV}^2$,
$\Delta m^2_{31} = 3\cdot 10^{-3} ~ {\rm eV}^2$,
$|V_{e1}| = 0.816$, $|V_{e2}| = 0.571$,
$|V_{\mu 3}| = 0.640$, and $J = \pm 0.020$  
have typically been input.}
\end{figure}

\begin{figure}[t]
\vspace{-0.8cm}
\epsfig{file=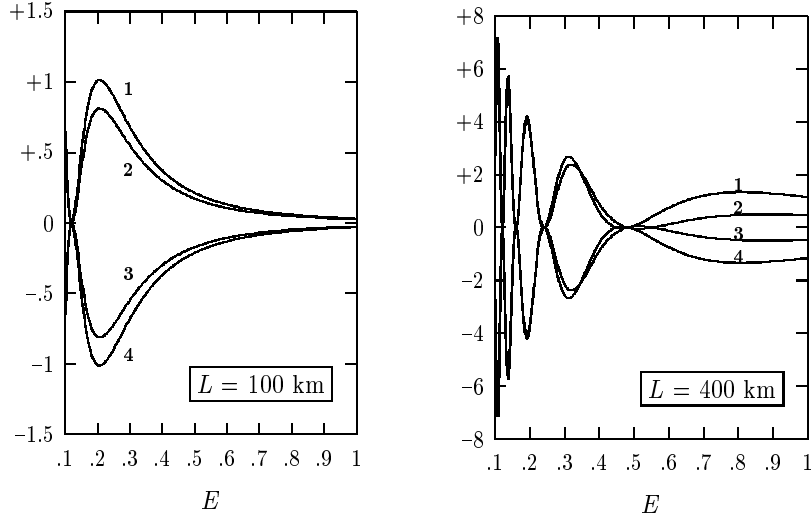,bbllx=1cm,bblly=6cm,bburx=19cm,bbury=28cm,%
width=15cm,height=20cm,angle=0,clip=}
\vspace{-12.2cm}
\caption{\small Dependence of the CP-violating asymmetry 
$\Delta {\bf P}$ (in unit of $10^{-2}$)
on the beam energy $E$ (in unit of GeV) and on the signs of
$(\Delta m^2_{21}, \Delta m^2_{31}$) -- {\bf 1}: ($-$, $+$);
{\bf 2}: ($-$, $-$); {\bf 3}: ($+$, $+$); {\bf 4}: ($+$, $-$).
Here $|\Delta m^2_{21}| = 5\cdot 10^{-5} ~ {\rm eV}^2$,
$|\Delta m^2_{31}| = 3\cdot 10^{-3} ~ {\rm eV}^2$,
$|V_{e1}| = 0.816$, $|V_{e2}| = 0.571$,
$|V_{\mu 3}| = 0.640$, and $J = \pm 0.020$  
have typically been input.}
\end{figure}

\end{document}